\documentstyle[preprint,aps,epsfig]{revtex}
\tighten
\begin{document}
\title{Relationship Between the Kobayashi-Maskawa and Chau-Keung       
       Presentations of the Quark Mixing Matrix}
\author{Valentine~E.~Kuznetsov\protect\thanks{Permanent address:
        Laboratory of Theoretical Physics, Irkutsk State University,
        Irkutsk 664003, Russia.}}
\address{Joint Institute for Nuclear Research, Dubna 141980, Russia}
\author{Vadim~A.~Naumov$^*$}
\address{INFN, Sezione di Firenze, Firenze I-50125, Italy} 
\date{\today}
\maketitle
\begin{abstract}
We discuss the formulas for one-to-one correspondence between the two
popular parametrizations of the quark mixing matrix and the confidence
limits for the mixing parameters.
\end{abstract}
\pacs{12.15.Ff, 12.15.Hh}
\narrowtext

After the pioneering work by Kobayashi and Maskawa \cite{Kobayashi73}
different presentations for the quark mixing matrix
$$ {\bf V} = 
\left(\begin{array}{ccc} V_{ud} & V_{us} & V_{ub} \\
                         V_{cd} & V_{cs} & V_{cb} \\
                         V_{td} & V_{ts} & V_{tb} \end{array}\right)$$
have been proposed (see Ref.~\cite{PDG94} and references therein).
It is however well known that no physics can depend on the
parametrization of the mixing matrix and its specific form is mostly a
matter of taste. It is unlikely that one can find the form of the
mixing matrix best suited to all physical problems and so it is useful
to have the exact formulas for one-to-one correspondence between
different forms in hand. In this Brief Report we give, as a case in
point, the formulas for the two popular parametrizations: the
``standard'' and ``the'' Kobayashi-Maskawa ones.

The ``standard'' form of the flavor mixing matrix advocated by the
Particle Data Group \cite{PDG94} following Harari and
Leurer~\cite{Harari86}, is that of Chau and Keung~\cite{Chau84}: 
$$ {\bf V}^{\rm (CK)} = \left(\begin{array}{ccc}
 c_{12}c_{13}                                    &
 s_{12}c_{13}                                    &
 s_{13}e^{-i\delta_{13}}                           \\
-s_{12}c_{23}-c_{12}s_{23}s_{13}e^{i\delta_{13}} &
 c_{12}c_{23}-s_{12}s_{23}s_{13}e^{i\delta_{13}} &
 s_{23}c_{13}                                      \\
 s_{12}s_{23}-c_{12}c_{23}s_{13}e^{i\delta_{13}} &
-c_{12}s_{23}-s_{12}c_{23}s_{13}e^{i\delta_{13}} &
 c_{23}c_{13}                                   \end{array}\right), $$
where $s_{jk} = \sin\theta_{jk}$ and $c_{jk} = \cos\theta_{jk}$ for
$j,k = 1,2,3$; the mixing angles $\theta_{jk}$ lie in the first
quadrant and $0 \leq \delta_{13} < 2\pi$.

For the Kobayashi-Maskawa (KM) parametrization \cite{Kobayashi73} we
will use the following form~\cite{footnote1}:
$$ {\bf V}^{\rm (KM)} = \left(\begin{array}{ccc}
 c_1    & s_1c_3                      & s_1s_3                      \\
-s_1c_2 & c_1c_2c_3-s_2s_3e^{i\delta} & c_1c_2s_3+s_2c_3e^{i\delta} \\
-s_1s_2 & c_1s_2c_3+c_2s_3e^{i\delta} & c_1s_2s_3-c_2c_3e^{i\delta}
                                                \end{array}\right), $$
where $s_i = \sin\theta_i$ and $c_i = \cos\theta_i$ for $i = 1,2,3$;
the mixing angles $\theta_i$ lie in the first quadrant and
$-\pi < \delta \leq \pi$.

The charged weak currents of the standard model
$$ \overline{U}\gamma_\mu\left(\frac{1+\gamma_5}{2}\right){\bf V}D  $$
are invariant under the gaugelike transformations of the mass
eigenstates $U = (u,c,t)^T$ and $D = (d,s,b)^T$ and of the mixing
matrix ${\bf V}$
\begin{mathletters}                                      \label{1}
\begin{equation}
U \rightarrow \exp(i{\bf\Omega}_U)U, \quad
D \rightarrow \exp(i{\bf\Omega}_D)D,                     \label{1a}
\end{equation}
\begin{equation}
{\bf V} \rightarrow {\bf V}' =
     \exp(i{\bf\Omega}_U){\bf V}\exp(-i{\bf\Omega}_D ),  \label{1b} 
\end{equation}
\end{mathletters}
with arbitrary real diagonal matrices ${\bf\Omega}_U$ and
${\bf\Omega}_D$.
Thus the matrices ${\bf V}'$ and ${\bf V}$ are equivalent.

The following identities also hold true:
\begin{equation}
{\rm tr}\,({\bf\Omega}_U-{\bf\Omega}_D) =
       \arg\left(\det{\bf V}'\det{\bf V}^\dagger\right), \label{2}
\end{equation}
\begin{equation}
V_{\alpha i}'V_{\beta j}'\left(V_{\alpha j}'V_{\beta i}'\right)^* = 
V_{\alpha i} V_{\beta j} \left(V_{\alpha j} V_{\beta i} \right)^*.
                                                          \label{3}
\end{equation}
{}From Eq.~(\ref{3}) in particular it follows that the measure of CP
violation
$$ J =  {\rm Im}\left(V_{\alpha i}V_{\beta j}
                      V_{\alpha j}^*V_{\beta i}^*\right)
     = -{\rm Im}\left(V_{\alpha i}V_{\beta j}V_{\gamma k}
                                         \det{\bf V}^\dagger\right )$$
(where the triplets $\alpha,\beta,\gamma$ and $i,j,k$ are arbitrary
cyclic permutations of the $u,c,t$ and $d,s,b$, respectively) is
invariant under the transformation (\ref{1b}).

The substitution
${\bf V} = {\bf V}^{\rm (KM)}$ and ${\bf V}' = {\bf V}^{\rm (CK)}$
gives the sought relationship. Using the identities
$|V_{\alpha i}^{\rm (CK)}| = |V_{\alpha i}^{\rm (KM)}|$, it is a
simple matter to derive the relations between the mixing angles
(cf.~\cite{Fritzsch85}):
$$ s_{12}=\frac{s_1c_3}{\sqrt{1-s_1^2s_3^2}}, \quad s_{13}=s_1s_3,  $$
$$ s_{23}=\sqrt{\frac{c_1^2c_2^2s_3^2+s_2^2c_3^2
                       +2c_1s_2c_2s_3c_3\cos\delta}{1-s_1^2s_3^2}}, $$
and from Eqs.~(\ref{1b}) and (\ref{3}), we find the relationship
between the CP violating phases $\delta_{13}$ and $\delta$:
\begin{eqnarray*}
\sin\delta_{13} &=& \frac{c_2s_2(1-s_1^2s_3^2)\sin\delta}{\Lambda}, \\
\cos\delta_{13} &=& \frac{\Lambda_ 0 }{\Lambda}\cos^2\frac{\delta}{2}
                   +\frac{\Lambda_\pi}{\Lambda}\sin^2\frac{\delta}{2}.
\end{eqnarray*}
Here
\begin{eqnarray*}
\Lambda     &\equiv& \Lambda(\theta_1,\theta_2,\theta_3;\delta)
               =   (1-s_1^2s_3^2)c_{23}s_{23}                       \\
            &  =   & \sqrt{\Lambda_ 0 ^2\cos^2\frac{\delta}{2} 
                          +\Lambda_\pi^2\sin^2\frac{\delta}{2}
                          +4c_1^2s_2^2c_2^2s_3^2c_3^2\sin^2\delta}, \\
\Lambda_0   &  =   & (c_2c_3-c_1s_2s_3)(c_1c_2s_3+s_2c_3),          \\
\Lambda_\pi &  =   & (c_2c_3+c_1s_2s_3)(c_1c_2s_3-s_2c_3),       
\end{eqnarray*}
and it is anticipated that $\Lambda \neq 0$ in conformity with the
experimental limits~\cite{footnote2}.
Clearly $\Lambda = |\Lambda_\delta|$ at $\delta = 0,\pi$.

The implicit form of the phase matrices ${\bf\Omega}_U$ and
${\bf\Omega}_D$ is defined modulo a common matrix
$c\cdot{\rm diag}(1,1,1)$ with $c$ an arbitrary constant.
This constant may be chosen so that the phases of the fields
$u$ and $d$ do not change under the transformation (\ref{1a}).
By direct substitution, one can verify that
\begin{eqnarray*}
{\bf\Omega}_U &=&{\rm diag}\left(0,\,
                   \frac{\delta_{13}-\delta+\Omega}{2},\,
                   \frac{\delta_{13}-\delta+2\pi-\Omega}{2}\right), \\
{\bf\Omega}_D &=&{\rm diag}\left(0,\,0,\,\delta_{13}\right),            
\end{eqnarray*}
where the angle $\Omega$ is determined by 
\begin{eqnarray*}
 \sin\Omega  &=& \frac{c_1s_3c_3\sin\delta}{\Lambda},               \\
 \cos\Omega  &=& \frac{\Lambda_ 0 }{\Lambda}\cos^2
\frac{\delta}{2}-\frac{\Lambda_\pi}{\Lambda}\sin^2\frac{\delta}{2}. 
\end{eqnarray*}

It is easy to verify that $\sin\delta_{13}=\sin\Omega=0$ and
$\cos\delta_{13}=(-1)^n\cos\Omega=\mbox{sign}\,(\Lambda_\delta)$ at
$\delta = n\pi$, $n = 0,1$. Or put in another way,
$$ \delta_{13}\Bigr|_{\delta =  0 }=    \Omega\Bigr|_{\delta=0}  =
 \left\{\begin{array}{lcl}   0, & \mbox{if} & c_2c_3 > c_1s_2s_3, \\
                           \pi, & \mbox{if} & c_2c_3 < c_1s_2s_3, 
                                                 \end{array}\right. $$
and
$$ \delta_{13}\Bigr|_{\delta = \pi}=\pi-\Omega\Bigr|_{\delta=\pi}=
 \left\{\begin{array}{lcl}   0, & \mbox{if} & s_2c_3 < c_1c_2s_3, \\
                           \pi, & \mbox{if} & s_2c_3 > c_1c_2s_3.
                                                 \end{array}\right. $$

The formulas for the inverse transformation may be obtained from the
foregoing ones by interchanging
\begin{eqnarray*}
s_1 \leftrightarrow c_{13}, \quad s_2 \leftrightarrow s_{23}, \quad
s_3 \leftrightarrow c_{12},                                   \\
c_1 \leftrightarrow s_{13}, \quad c_2 \leftrightarrow c_{23}, \quad
c_3 \leftrightarrow s_{12},                                   \\
\sin\delta \leftrightarrow  \sin\delta_{13},                  \quad
\cos\delta \leftrightarrow -\cos\delta_{13}.                         
\end{eqnarray*}

\begin{figure}\mbox{\epsfig{file=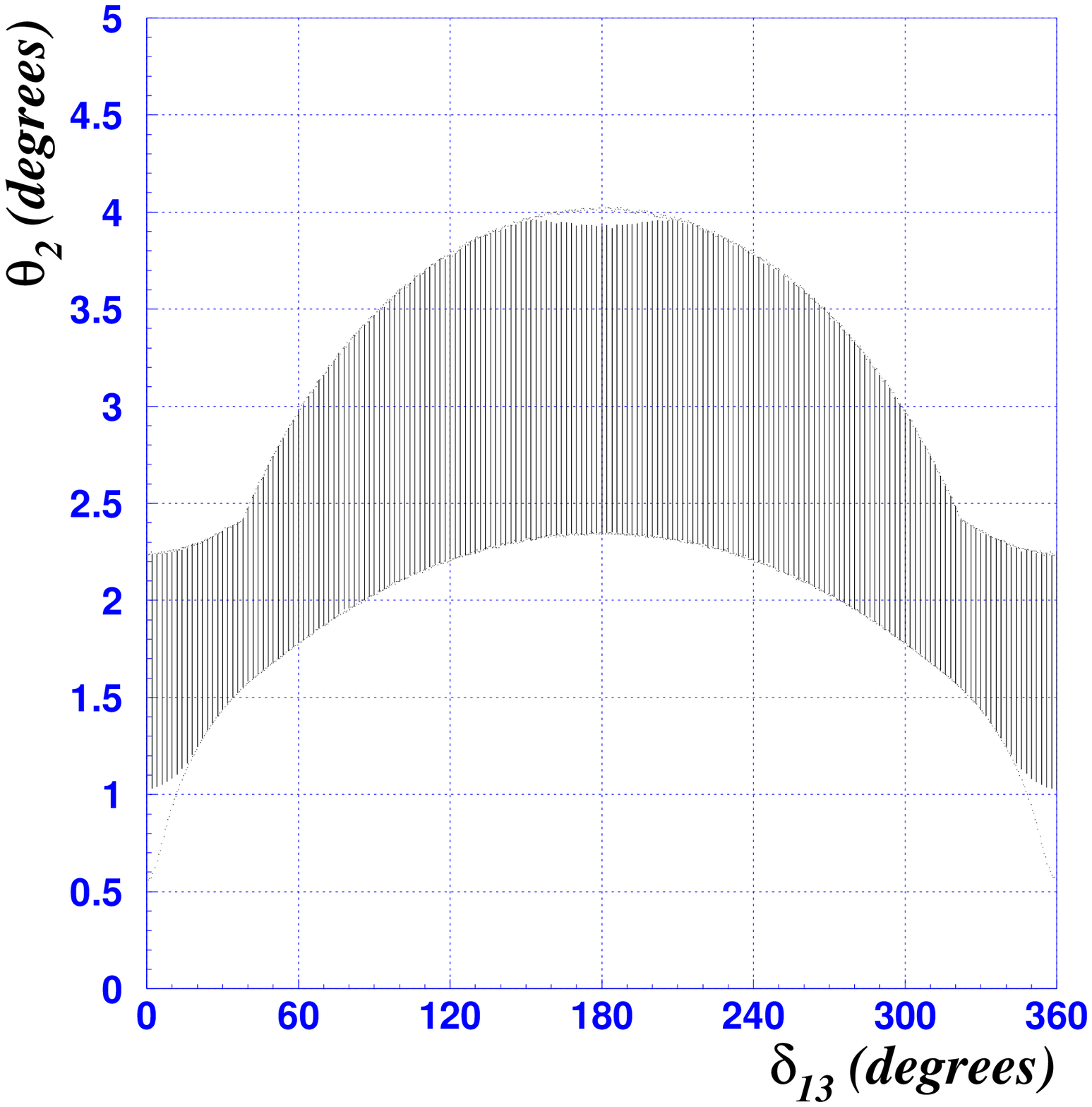,width=15.5cm}}
\protect\caption{\label{Fig1}
        $\theta_2$ vs $\delta_{13}$ for the CK mixing angles
        $\theta_{jk}$ from the 90\% confidence intervals given by
        $s_{12}=0.218$ to 0.224, $s_{23}=0.032$ to 0.048, and
        $s_{13}=0.002$ to 0.005 \protect\cite{PDG94}. The shaded area
        satisfies the unitarity constraints.}
\end{figure}

With the derived formulas we can obtain the 90\% confidence limits for
the KM mixing angles using the experimental constraints on the CK
mixing angles and the constraints on the magnitude of the elements
$V_{\alpha i}$ imposed by unitarity~\cite{PDG94}. The angle $\theta_1$
lies in the narrow interval $12.6^\circ \div 12.9^\circ$
($s_1 = 0.218$ to 0.224) while the confidence interval for the angle
$\theta_3$ proves to be comparatively wide:
$0.512^\circ \div 1.31^\circ$ ($s_3 = 0.009$ to 0.023).
Fig.~\ref{Fig1} shows the angle $\theta_2$ vs $\delta_{13}$.
The maximum uncertainty in this angle is about
$1.6^\circ$ at $\delta_{13} \sim 160^\circ - 200^\circ$.

\begin{figure}\mbox{\epsfig{file=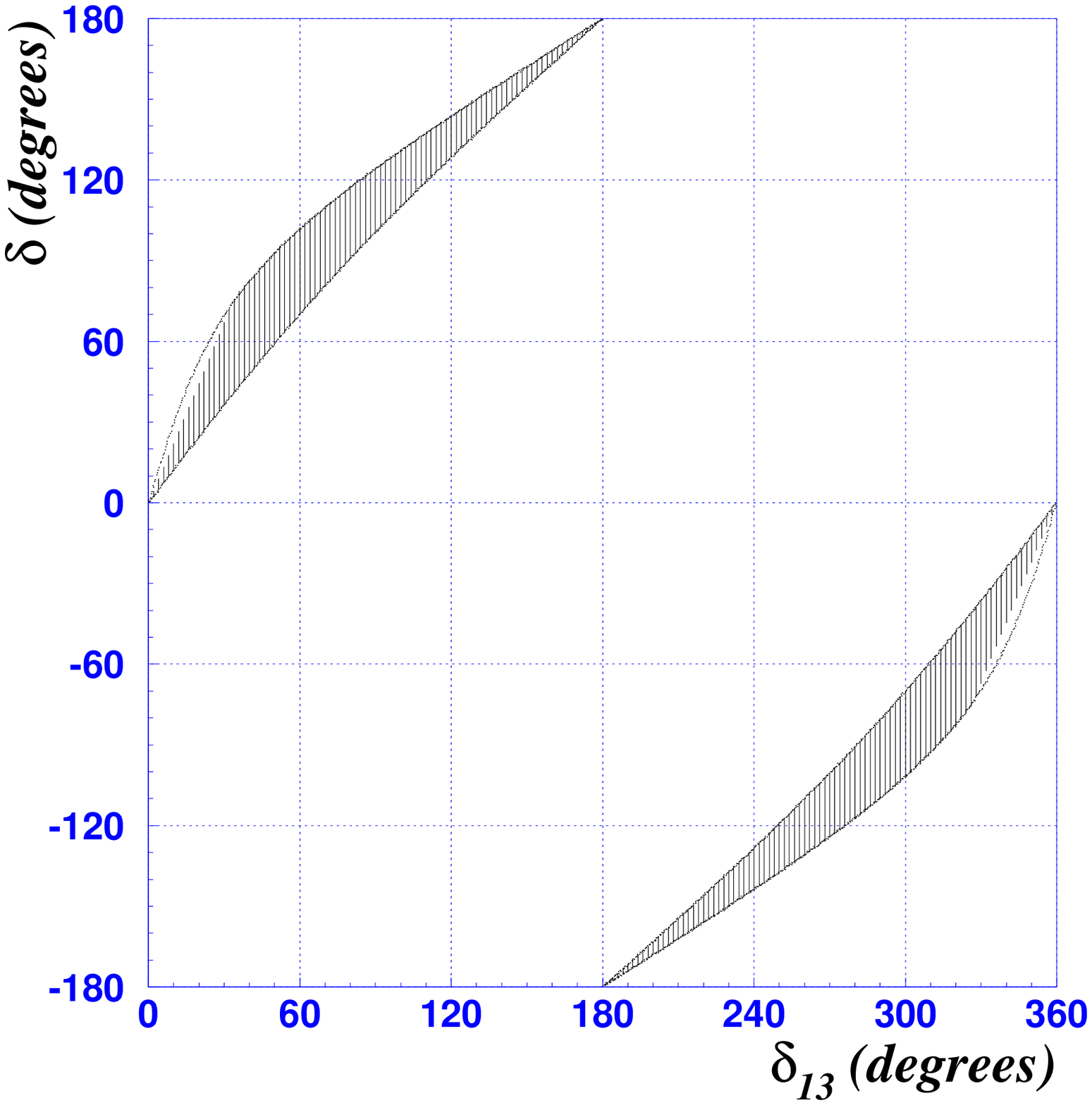,width=15.5cm}}
\protect\caption{\label{Fig2}
        $\delta$ vs $\delta_{13}$ for the same intervals of the CK
        mixing angles as in Fig.~\protect\ref{Fig1}. The shaded areas
        satisfy the unitarity constraints.}
\end{figure}

In Fig.~\ref{Fig2} we show the function $\delta$ vs $\delta_{13}$ for
the same range of the CK mixing angles. One can see that with the
modern 90\% confidence limits on the $|V_{\alpha i}|$, the maximum
uncertainty in prediction of the KM phase $\delta$ at a given value of
the CK phase $\delta_{13}$ is about $34^\circ$. Fig.~\ref{Fig2}
demonstrates that the current 90\% limits on the mixing angles present
unambiguous correspondence: $\delta = 0$ at $\delta_{13} = 0$ and
$\delta = \pi$ (or $-\pi$, that is the same) at $\delta_{13} = \pi$.
It is also seen that the maximal CP nonconservation in the standard
presentation ($\delta_{13} = \pi/2$ or $3\pi/2$) leads to $\delta$
different from $\pm\pi/2$ \cite{footnote3}. Namely at
$\delta_{13} = \pi/2$ the phase $\delta$ lies in the interval
$1.75 \div 2.18$ ($100.3^\circ\div124.8^\circ$). The measure of CP
nonconservation, $J$, is of course independent on the parametrization:
\begin{eqnarray*}
J &=& \frac{1}{8}\sin\delta\sin\theta_1\prod_i    \sin2\theta_i    \\
  &=& \frac{1}{8}\sin\delta_{13}\cos\theta_{13}
                                       \prod_{j<k}\sin2\theta_{jk} \\
  &=& (1.36 \div 5.23) \times 10^{-5}\sin\delta_{13}.
\end{eqnarray*}

Summarizing we derived the exact formulas connecting the KM and CK
(standard) presentations of the mixing matrix which may be helpful for
the study of the CP violation in quark sector.

\acknowledgments

We are grateful to the PPE division of CERN, where this paper was
made, for its hospitality.

\newpage
\newpage

\begin{references}
\bibitem{footnote2} The equality $\Lambda = 0$ only holds for the
         pure two-quark mixing (if $c_{23}s_{23} = 0$) or in the
         absence of any mixing at all (if $c_1 = c_3 = 0$).
\end{references}
\end{document}